\documentclass[aps,prd,nofootinbib,preprintnumbers,superscriptaddress,12pt]{revtex4-1} 
\pdfoutput=1
\usepackage{amsmath,amssymb,bm,comment}
\usepackage{graphicx}
\usepackage{color}
\usepackage[pdftex]{hyperref}
\usepackage[utf8]{inputenc}

\def\MK#1{\textcolor{blue}{#1}}

\newcommand{\B}{{B}}
\newcommand{\F}{\mathcal{F}}
\renewcommand{\O}{\mathcal{O}}
\newcommand{\V}{\mathcal{V}}
\newcommand{\Z}{\mathcal{Z}}
\newcommand*{\half}{\tfrac{1}{2}}
\newcommand{\FQFT}{f_{\rm QFT}}
\newcommand{\SQFT}{S_{\rm QFT}}
\newcommand{\FEM}{f_{\rm EM}}
\newcommand{\SEM}{S_{\rm EM}}

\newcommand{\Nfour}{\mathcal{N}\,{=}\,4}
\newcommand{\Nc}{N_{\rm c}}
\newcommand{\tr}{\mathop{\rm tr}}
\newcommand{\LambdaH}{\Lambda_{\rm H}}

\begin{document}




\title		{Universal magnetoresponse in QCD and $\Nfour$ SYM}

\author		{Gergely~Endr\H{o}di}
\email		{endrodi@th.physik.uni-frankfurt.de}
\affiliation	{Institute for Theoretical Physics,
		Goethe Universit\"at Frankfurt,
		D-60438 Frankfurt am Main, Germany}
\author		{Matthias~Kaminski}
\email		{mski@ua.edu}
\affiliation	{Department of Physics and Astronomy,
		University of Alabama, Tuscaloosa, AL 35487, USA}
\author		{Andreas~Sch\"afer}
\email		{andreas.schaefer@physik.uni-regensburg.de}
\affiliation	{Institut f\"ur Theoretische Physik,
		Universit\"at Regensburg,
		93040 Regensburg, Germany}
\author		{Jackson~Wu}
\email		{jmwu@ua.edu}
\affiliation	{Department of Physics and Astronomy,
		University of Alabama, Tuscaloosa, AL 35487, USA}
\author		{Laurence~Yaffe}
\email		{yaffe@phys.washington.edu}
\affiliation	{Department of Physics, University of Washington,
		Seattle, WA 98195-1560, USA}



\begin{abstract}
    Using recent lattice data on the thermodynamics of QCD
    in the presence of a background magnetic field,
    we show that the ratio of transverse to longitudinal pressure
    exhibits, to good accuracy, a simple scaling behavior
    over a wide range of temperature and magnetic field,
    essentially depending only on the ratio $T/\sqrt B$.
    We compare this QCD response to the corresponding
    magnetoresponse in maximally supersymmetric Yang Mills theory.
    Given suitable calibrations defining the comparison,
    we find excellent agreement.
    This may be viewed as a further test of the applicability
    of holographic models for hot QCD.
\end{abstract}

\maketitle


\tableofcontents

\newpage
\section{Introduction}

    Gauge-gravity duality has enabled quantitative studies of the
dynamics of certain strongly coupled non-Abelian plasmas
\cite{Witten:1998zw,
Aharony:1999ti,
DeWolfe:2013cua,
Adams:2012th}.
Despite the limitations of holographic models (involving large $\Nc$,
strong coupling limits, and supersymmetry),
they have provided important insight into key properties of
quark-gluon plasma as observed in relativistic heavy-ion collisions,
including fast ``thermalization'' and the applicability of near-ideal
hydrodynamics
\cite{Chesler:2010bi,
Chesler:2008hg,
Janik:2005zt,
Chesler:2009cy,
Chesler:2015wra}.

In this paper, we compare the magnetoresponse of QCD plasma and
maximally supersymmetric Yang Mills ($\Nfour$ SYM) plasma ---
the non-Abelian plasma with the simplest holographic description.
Specifically, we examine the change in thermodynamic properties
induced by a homogeneous background magnetic field.
The response to an applied magnetic field is a useful
probe of the dynamics in many condensed matter systems.
In our context, an examination of magnetoresponse is also
motivated by work suggesting that electromagnetic fields
(albeit transient) may have significant effects in heavy ion collisions
\cite{Kharzeev:2004ey,
Kharzeev:2012ph}.

Holographic models have been found to describe rather accurately
many aspects of strongly coupled QCD dynamics, despite the fact that
QCD is neither conformal, supersymmetric, nor infinitely strongly coupled.
Any reasonable measure of QCD coupling strength in experimentally accessible
quark-gluon plasma is order unity, far from the infinite coupling limit,
and $\Nc=3$ appears equally far from $N_c = \infty$.
The apparent robustness of AdS/CFT predictions,
despite these limitations,
has prompted numerous investigations.
It has been shown, for example,
that finite coupling corrections
in many thermal quantities are modest
\cite{Waeber:2015oka,Waeber:2018bea},
and that $\Nc$ dependence
is essentially trivial, with extensive quantities
simply scaling with the number of gauge fields
\cite{Panero:2009tv,Bali:2013kia}.

To investigate whether a similar robustness exists with respect to conformal
symmetry it is natural to examine the effects of deformations 
which explicitly break conformal symmetry.
Adding a background magnetic field is such a non-conformal
deformation.

In the presence of an external electromagnetic field,
the definition of the QCD contribution to the total stress-energy tensor
depends on a choice of renormalization point.
This issue is discussed in section~\ref{sec:theoriesMethods},
which reviews basic properties of the stress-energy tensor of a
quantum field theory when minimally coupled to a
non-dynamical electromagnetic field.

\begin{figure}[t]
\centering
\hspace{-20pt}%
\includegraphics[height=2.6in]{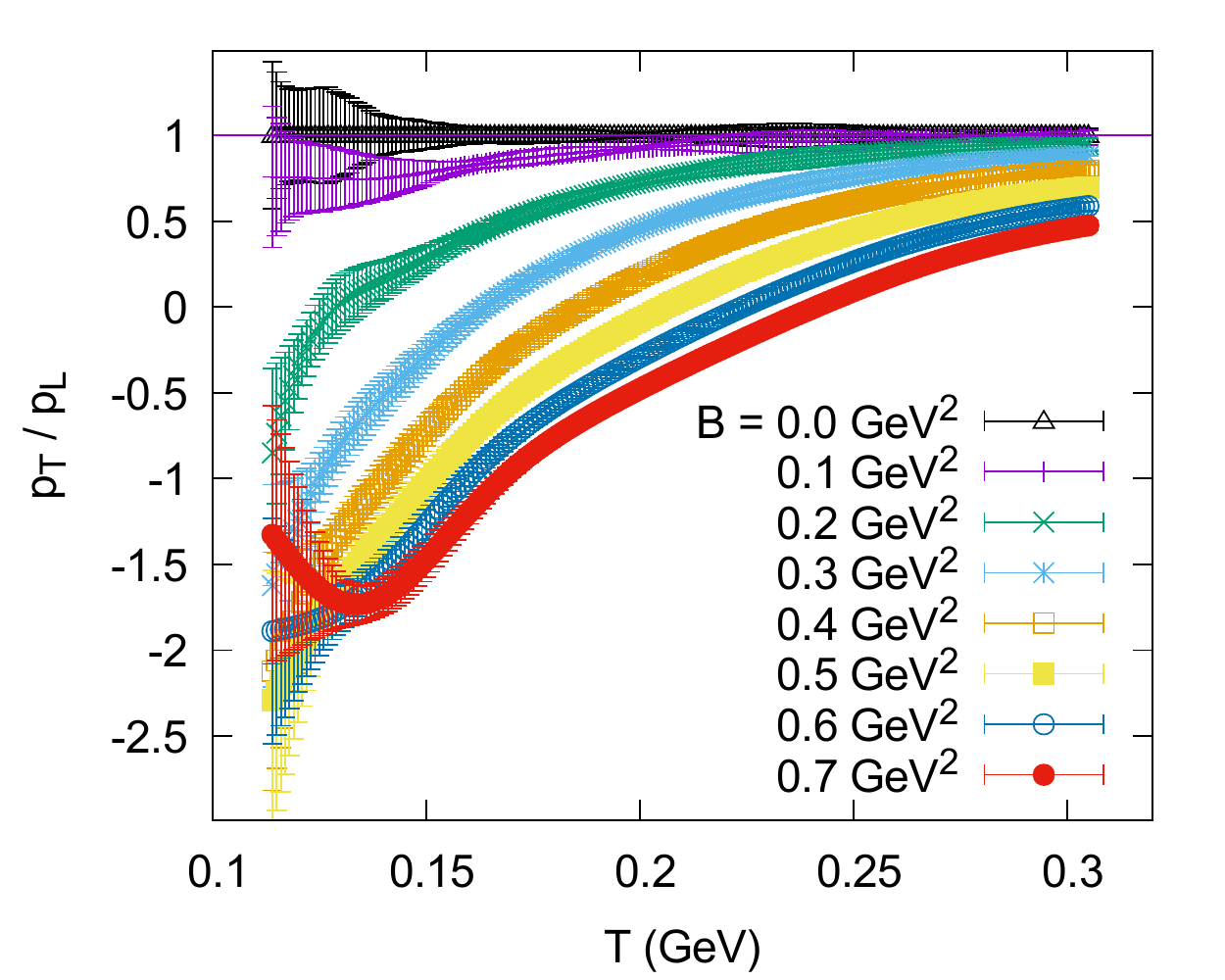}%
\includegraphics[height=2.6in]{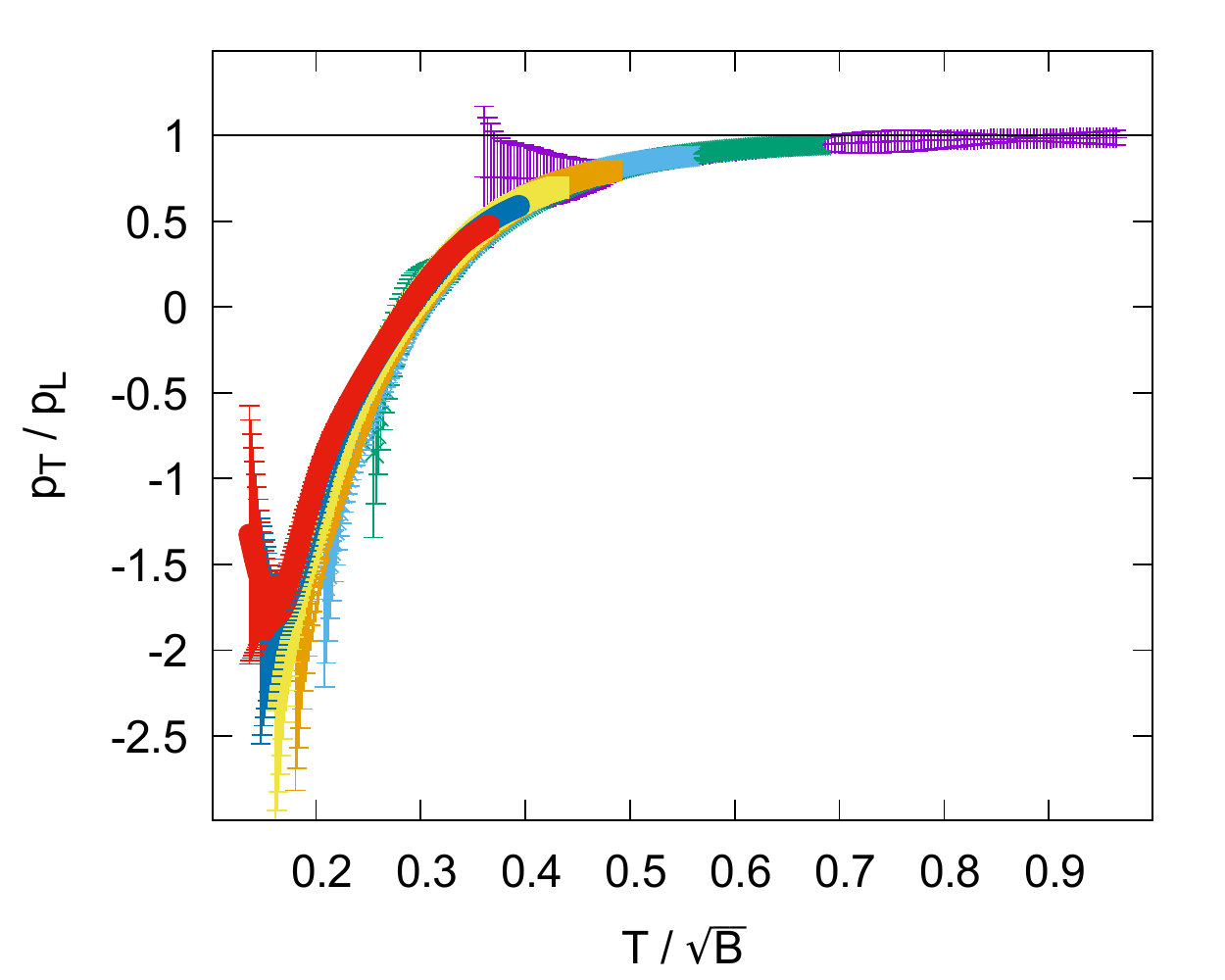}%
\hspace{-20pt}
\vspace*{-10pt}
\caption
    {%
    The ratio $R \equiv p_T/p_L$ of transverse to longitudinal pressure
    in QCD,
    defined with renormalization point $\mu=\LambdaH$,
    for various values of external magnetic field $B$.
    (See Section~\ref{sec:latticeQCD} for details.)
    Left panel: $R$ plotted as a function of $T$.
    Right panel: $R$ plotted as a function of $T/\sqrt{B}$.
    The different colors indicate different values of 
    the magnetic field and are identical in the two panels.
    }
\label{fig:scaling}
\end{figure}

For QCD magnetoresponse,
we take as input results from recent high quality lattice gauge theory
calculations of QCD thermodynamics for a wide range of 
background magnetic field $B$ and
temperature $T$~\cite{Bali:2013esa,Bali:2014kia}.
A natural measure of anisotropy in the system is
the transverse to longitudinal pressure ratio $p_T/p_L$,
shown in the left panel of Fig.~\ref{fig:scaling} as a function of 
$T$ for different magnetic fields. 
Interestingly, we find that this ratio
exhibits, to good accuracy, a simple scaling behavior over
a wide range of temperature and magnetic field.
As shown in the right panel of the same figure, when
plotted as a function of $T/\sqrt B$ data from widely differing
values of $T$ and $B$ essentially collapse onto a single curve.
The underlying lattice QCD data is discussed
in more detail in Section~\ref{sec:latticeQCD},
(Deviations from this scaling behavior appear to be present
at the lowest temperatures and highest magnetic fields,
but the growth of the error bars precludes making any
definitive statement about this region.)

\begin{figure}[htp]
\includegraphics[height=3.0in]{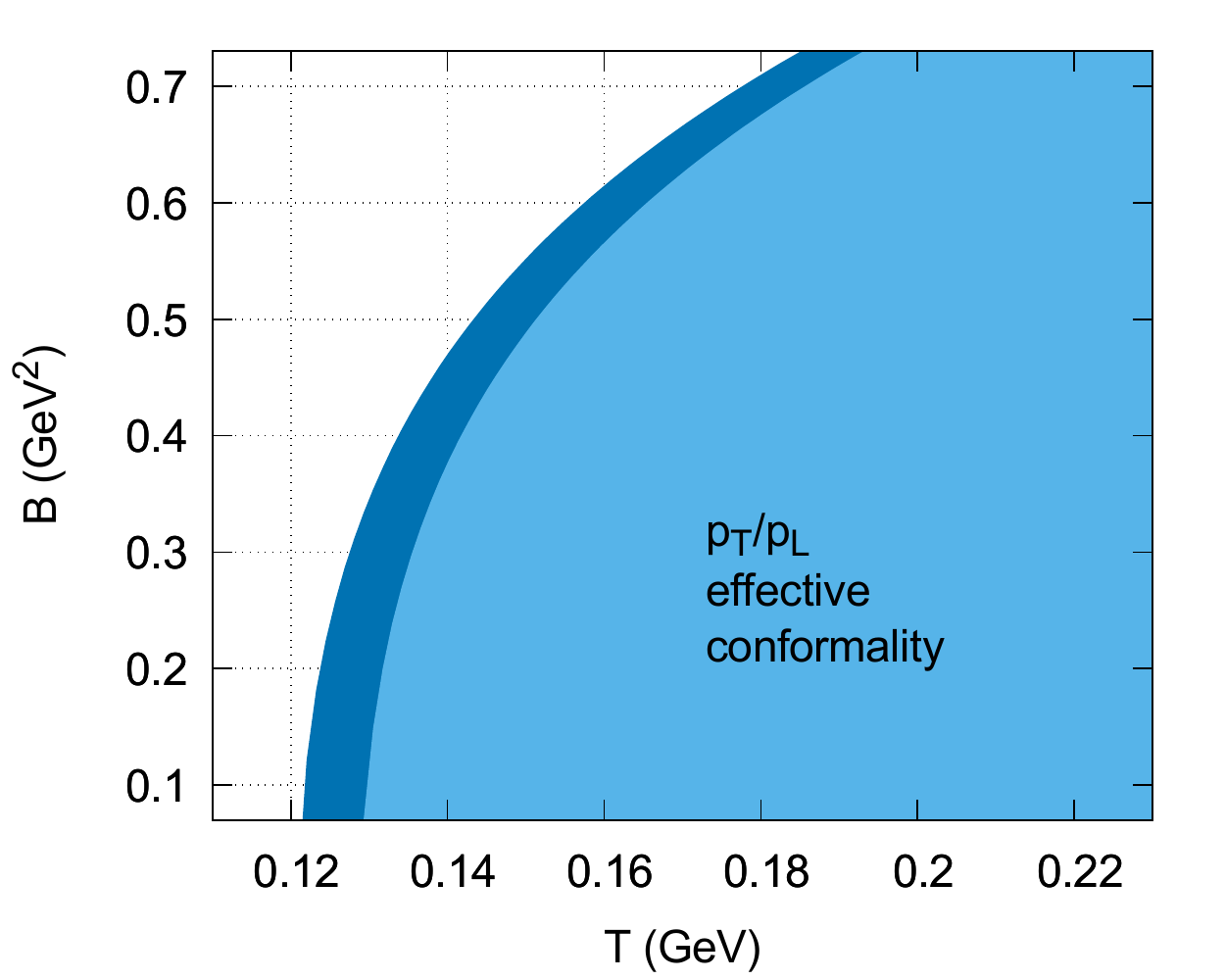}
\vspace*{-10pt}
\caption
    {%
    Region of effective conformality, in the plane of
    temperature and magnetic field,
    of the QCD pressure anisotropy ratio $p_T/p_L$.
    Within the blue region
    QCD and $\Nfour$ SYM, appropriately compared,
    give identical values for this ratio to within the
    error estimates of the lattice data.
    The dark blue band indicates the uncertainty in the border of
    this region arising from lattice errors.
    }
\label{fig:optimizedPressureRatioDiag}
\end{figure}

In any conformal field theory,%
\footnote
    {%
    See Ref.~\cite{Nakayama:2013is} for a careful discussion
    of the relation between scale invariance
    and full conformal symmetry.
    }
the lack of intrinsic scales
automatically implies that the magnetoresponse (appropriately defined)
can only depend on the dimensionless ratio $T/\sqrt B$.
So, having found near-universal scaling behavior in the QCD magnetoresponse, 
it is natural to compare this response to that of the simplest
holographic model, namely conformal $\Nfour$ SYM in the strong coupling
and large $N$ limit, for which the dual description reduces to 5D
Einstein-Maxwell theory.
We briefly review $\Nfour$ SYM theory and its coupling to a background
EM field in section~\ref{sec:SYM},
and then compare the QCD and $\Nfour$ SYM
magnetoresponse in Section~\ref{sec:comparison}.
A key issue, discussed in some detail, concerns how best to make
such a comparison given the unavoidable renormalization point dependence
of the quantities under study.

For impatient readers, our end result is shown graphically in 
Fig.~\ref{fig:optimizedPressureRatioDiag}.
Within the shaded region of the temperature-magnetic field plane,
the QCD and $\Nfour$ SYM results for the pressure anisotropy ratio $p_T/p_L$,
appropriately compared, are found to be identical to within
the errors of the lattice data.
Our final section~\ref{sec:discussion} discusses the implications of
this result and possible future directions.
An appendix contains a few details regarding a high temperature matching
procedure for relating QCD and SYM quantities.

\section{Quantum field theory with an external magnetic field}\label{sec:theoriesMethods}

\subsection{Thermodynamics}\label{sec:EoS}

Consider a quantum field theory (QFT) 
minimally coupled to an external electromagnetic $U(1)$~gauge
field $A_\mu^{\rm ext}$, with field strength
$
    F_{\mu\nu}^{\mathrm{ext}} =
    \partial_\mu A^{\mathrm{ext}}_\nu - \partial_\nu A^{\mathrm{ext}}_\mu
$.
A constant magnetic field $\mathbf{B}=B \, \mathbf{e}_z$ may be described
by the standard choice
$
    A_\mu^{\mathrm{ext}} = \frac{1}{2} B (x^1\delta^2_\mu - x^2 \delta^1_\mu)
$.
The total action of the theory may be written in the form
\begin{equation} \label{eq:QFTaction}
    S = \SQFT(B) + \SEM(e,B)\,,
    \qquad
    S_{\rm EM}(e,B) \equiv -\int d^4 x \> \frac{B^2}{2e^2} \,,
\end{equation}
where $\SEM$ is the classical Maxwell action,
specialized to a pure magnetic field,
with $e$ the (bare) electromagnetic coupling constant,
and the QFT action $\SQFT(B)$ includes the minimal coupling to the
background EM field.%
\footnote
    {%
    In this section, we work in Minkowski space with
    metric $\eta_{\mu\nu} = \mathrm{diag}(-1,+1,+1,+1)$.
    }
Here and henceforth, we choose to scale the
external gauge field $A_\mu^{\rm ext}$ so that the electromagnetic coupling $e$
does not appear in covariant derivatives, but instead $e^2$ is an inverse
factor in the Maxwell action.
Consequently, our $B$ is the same as $eB$ if a
conventional perturbative scaling of the $U(1)$ field is used.%
\footnote
    {%
    In particular, our magnetic field $B$ corresponds to
    $eB$ in Ref.~\cite{Bali:2013esa}.
    }

From the action~\eqref{eq:QFTaction} (generalized to curved space)
one derives the stress-energy tensor,
\begin{equation}
    T^{\mu\nu} =
    -2
    \frac{\delta S}{\delta g_{\mu\nu}} 
    = T^{\mu\nu}_{\mathrm{QFT}} 
    + T^{\mu\nu}_{\mathrm{EM}} \, ,
\label{eq:Tmunudef}
\end{equation}
where
$
    T^{\mu\nu}_{\mathrm{QFT}}=-2\delta \SQFT/\delta g_{\mu\nu}
$
is the QFT contribution to the total stress-energy tensor, while
\begin{equation}
    T^{\mu\nu}_{\mathrm{EM}}
    =
    -2 \frac{\delta S_{\rm EM}}{\delta g_{\mu\nu}}
    =
    \frac 1{e^2}
    \left(
	F^{\mu\alpha} F^{\nu\beta} \, \eta_{\alpha\beta}
	- \tfrac 14 \, \eta^{\mu\nu} F^{\alpha\beta} F_{\alpha\beta}
    \right) ,
\end{equation}
is the standard Maxwell stress-energy tensor (in Minkowski space).
Specialized to a constant magnetic field in the $z$-direction,
$
    T^{\mu\nu}_{\mathrm{EM}}
    =
    \frac{B^2}{2e^2} \>
    \mathrm{diag} (+1,+1,+1,-1)
$.

In a homogeneous equilibrium state, viewed in a rest frame
(with vanishing momentum density) aligned with the magnetic field,
diagonal elements of the expectation value
of the stress-energy tensor can be interpreted
as the proper energy density $\epsilon$, and pressures (or diagonal stresses) 
$p_x,\, p_y,\, p_z$
along the $x$-, $y$-, $z$-directions, respectively,
\begin{equation}
    \langle T^{\mu\nu}\rangle
    = 
    \begin{pmatrix}
    \;\epsilon\; & 0 & 0 & 0\\
    0 & p_x & 0 & 0 \\
    0 & 0 & p_y & 0 \\
    0 & 0 & 0 & p_z 
    \end{pmatrix} .
\label{eq:pressures}
\end{equation} 
The magnetic field defines a preferred direction in space,
and induces an anisotropy between the longitudinal pressure
$p_L \equiv p_z$ and
transverse pressure $p_T \equiv p_x = p_y$.
(Rotational symmetry about the magnetic field direction implies
that $p_x = p_y$.)
The relation between the pressure(s) and energy density 
constitutes the equation of state of the system.

Similarly to the action~\eqref{eq:QFTaction},
the thermodynamic grand potential (or Landau free energy),
\begin{equation}
    \F \equiv - {T \, \ln \Z } \,,
\end{equation}
may be separated into QFT and EM contributions,
\begin{equation}
    \F = \F_{\rm QFT}(B) + \F_{\rm EM}(e,B)\,,
    \quad\quad
    \F_{\rm EM}(e,B) = \V \, \frac{B^2}{2e^2}\,,
 \label{eq:Fdef}
\end{equation}
with $\V = L_x L_y L_z$ the spatial volume.
This separation, by definition, places all the response to the applied
magnetic field in the QFT contribution to the free energy.
Let $\FQFT = \F_{\rm QFT}/\V$ and $\FEM = \F_{\rm EM}/\V$
denote the corresponding free energy densities.
Derivatives of the free energy density $\FQFT$ with respect to the temperature or 
magnetic field define the entropy density and magnetization, respectively,
\begin{equation}
    s= -\frac{\partial \FQFT}{\partial T},
    \qquad
    M=-\frac{\partial \FQFT}{\partial B}\,.
\end{equation}
Similarly, pressures are defined by the response of the system
to compression in a given direction,
\begin{equation}
    p_i = -\frac{L_i}{\V} \, \frac{\partial \F_{\rm QFT}}{\partial L_i} \,.
\label{eq:pdef}
\end{equation}
In the thermodynamic limit,
for homogeneous systems,
the pressure is normally just minus the free energy density,
since the free energy is extensive in the volume. 
But with a background magnetic field, 
one must specify what is to be held fixed in the partial derivative defining
the pressure (\ref{eq:pdef})~\cite{Bali:2013esa}. 
The microscopic definition (\ref{eq:pressures}) of pressures as
stress-energy eigenvalues corresponds to a thermodynamic definition
in which the effect of compression is evaluated at a fixed
magnetic flux $\Phi=B \, L_xL_y$.%
\footnote
    {%
    To see this, note that the metric variation defining the 
    stress-energy (\ref{eq:Tmunudef}) leaves unchanged the
    EM flux across any 2-surface, $\Phi = \int_\Sigma F$,
    since integration of a two-form is metric independent.
    }
Consequently, the longitudinal and transverse pressures differ,
\begin{equation}
    p_L = -\FQFT \,,\qquad
    p_T = p_L - M \cdot B \,.
\label{eq:anisotropy}
\end{equation}
The thermodynamic relation $\FQFT=\epsilon-Ts$ implies that the
entropy and energy densities $s$ and $\epsilon$ are related via
\begin{equation}
    s=\frac{\epsilon+p_L}{T}\,. 
\end{equation}
A final quantity of interest is the interaction measure $I$,
which is (minus) the trace of the stress-energy tensor 
\begin{equation}
    I = -\langle T^{\mu}{}_\mu \rangle
    = \epsilon-2p_T-p_L = \epsilon-3p_L+2M\cdot B \,.
\label{eq:I}
\end{equation}
In a CFT, such as $\Nfour$ SYM,
conformal symmetry implies that the stress-energy tensor is traceless.
However, adding an external magnetic field $B$ is a non-conformal
deformation of the theory and induces a non-zero trace,
and hence a nonzero interaction measure.
For $\Nfour$ SYM (coupled to the external field in the manner described
below in section \ref{sec:SYM}),
\begin{equation}\label{eq:SYMTraceAnomaly}
    T_\mu^\mu= - \frac{\Nc^2{-}1}{4\pi^2} \, B^2 \,.\hspace{1.2in}
    \mbox{[$\Nfour$ SYM]}
\end{equation} 
For an asymptotically free theory like QCD, the stress-energy trace 
has an intrinsic contribution from the running of the coupling
(and quark masses terms),
plus the additional contribution from the external magnetic field.
Neglecting quark masses,
\begin{equation}
    T_\mu^\mu=
    - \beta(g^{-2}) \, \tfrac 14 \tr G_{\mu\nu}^2
    - \widetilde\beta(e^{-2}) \, \tfrac 12 B^2 \,,\qquad \mbox{[QCD]}
\label{eq:QCDtraceanom}
\end{equation}
where $G_{\mu\nu}$ is the gluon field strength,
$
    \beta(g^{-2}) \equiv \mu \partial_\mu \, g^{-2}
    =
    9/(4\pi^2) + \O(g^2)
$
is the renormalization group $\beta$-function for the $SU(3)$ inverse gauge coupling
(with three quark flavors),
and
$
    \widetilde\beta(e^{-2}) \equiv \mu \partial_\mu \, e^{-2}
    =
    -1/(3\pi^2) + \O(g^2)
$
is the corresponding electromagnetic $\beta$-function arising from the
three light quark flavors of QCD.

\subsection{Renormalization}
\label{sec:renorm}

In interacting quantum field theories,
bare parameters of the action undergo multiplicative renormalization
which introduces dependence on an arbitrary renormalization point $\mu$.%
\footnote
    {%
    With our definition of the magnetic field $B$, the Ward-Takahashi
    identity shows that $B$ receives no wavefunction renormalization
    and is scale independent.
    }
In particular, the renormalized electromagnetic coupling $e^2$
acquires logarithmic scale dependence and
satisfies a QED-like renormalization group equation,
\begin{equation}
    \mu \frac {d}{d\mu} \, e^{-2}
    \equiv \widetilde \beta(e^{-2})
    = -2b_1 \times [1 + \O(g^2)] \,,
\label{eq:beta(e^2)}
\end{equation}
with positive coefficient $b_1$.
Explicitly, for QCD,
\begin{subequations}\label{eq:b1}%
\begin{equation}
    b_1
    = \frac {N_c}{12\pi^2} \sum_f \> q_f^2
    = \frac {1}{6\pi^2} 
    \,,\hspace*{1.2in} \mbox{[QCD]}
\label{eq:EMbeta_QCD}
\end{equation}
where $q_f$ denotes the electromagnetic
charge assignments (i.e., EM charges in units of $e$)
of each quark flavor, and the explicit final form is
specialized to three flavor QCD.
For $\Nfour$ SYM theory,
\begin{equation}
    b_1 = \frac{N_c^2{-}1}{24\pi^2}
    \left[ \sum_w \> q_w^2 + \tfrac{1}{2} \sum_s \> q_s^2\right],
    \qquad \mbox{[$\Nfour$ SYM]}
\label{eq:EMbeta_SYM}
\end{equation}
\end{subequations}
where the sums run over all 
charged Weyl fermions $w$ and charged scalars $s$
with $q_w$ and $q_s$ 
denoting the corresponding electromagnetic charge assignments.
If the electromagnetic field is regarded as classical
(so that EM quantum fluctuations are neglected)
then the higher order corrections in the $\beta$-function
(\ref{eq:beta(e^2)}) are independent of $e^2$.
For QCD in a background magnetic field, the EM $\beta$-function
has higher order corrections proportional to
the non-Abelian coupling $g^2$, while for
$\Nfour$ SYM (in a background field),
no higher order corrections appear in
the EM $\beta$-function~(\ref{eq:beta(e^2)}) due to a
supersymmetric non-renormalization theorem.
Neglecting any such higher order corrections,
the solution to the renormalization group equation~(\ref{eq:beta(e^2)})
shows the usual logarithmic scale dependence,
\begin{equation}
    e(\mu_1)^{-2}
    =
    e(\mu_2)^{-2}
    - 2 b_1 \ln (\mu_1/\mu_2) \,.
\label{eq:logrunning}
\end{equation}

Physical observables, like the total free energy,
are necessarily independent of the renormalization point $\mu$.
However, the separation~\eqref{eq:Fdef} of the free energy
into QFT and background EM contributions requires choosing
the scale at which to evaluate the coupling $e$ appearing in the
background EM contribution.
So this separation is more properly written as
\begin{equation}
    f = \FQFT(B,\mu) + \FEM(e(\mu),B)\,,
\label{eq:Fdef2}
\end{equation}
with the scale dependence of $f_{\rm QFT}(B,\mu)$ necessarily
canceling that of the EM term, so that
\begin{equation}
    \mu \frac d{d\mu} \, 
    \FQFT(B,\mu) = b_1 \, B^2 \,.
\label{eq:RG-f}
\end{equation}
Similarly, the QFT stress-energy tensor acquires scale dependence,
\begin{equation}
    \mu \frac {d}{d\mu} \, T^{\alpha\beta}_{\rm QFT}
    =
    2 b_1 \left(
	F^{\alpha\gamma} F^{\beta\delta}\eta_{\gamma\delta}
	-
	\tfrac 14 \, \eta^{\alpha\beta} F^{\gamma\delta} F_{\gamma\delta}
    \right) ,
\end{equation}
which precisely cancels the scale dependence of the EM stress-energy tensor.
Hence,
\begin{subequations}%
\label{eq:changeRenScale}
\begin{align}
    \epsilon(\mu')-\epsilon(\mu) &= b_1 \, B^2 \, \ln \frac {\mu'}{\mu} \,,
\\
    p_{T}(\mu')-p_{T}(\mu) &= b_1 \, B^2 \, \ln\frac{\mu'}{\mu} \,,
\\
    p_L(\mu')-p_L(\mu) &= -b_1 \, B^2 \, \ln\frac{\mu'}{\mu} \,.
\end{align}
\end{subequations}
This scale dependence induced by the separation of QFT response from the
background EM contributions is unavoidable, since the background field
contributions (for realistic values of the electromagnetic coupling) are
orders of magnitude larger than the matter-induced response~\cite{Bali:2014kia}
and would otherwise overshadow interesting features in the magnetic field
dependence of the QFT response.

\section{Lattice quantum chromodynamics} \label{sec:latticeQCD}

We consider QCD with $2+1$ flavors of dynamical quarks with physical masses.
The quarks have their usual electric charge assignments,
$q_u=+2/3$ and $q_d=q_s=-1/3$.
The action has the form \eqref{eq:QFTaction}, with 
$S_{\mathrm{QFT}}=S_{\mathrm{QCD}}$ and
covariant derivatives augmented to include the background
$U(1)_{\rm EM}$ gauge field.
The lattice regularized Euclidean functional integral was simulated
non-perturbatively using a staggered fermion discretization and
three different lattice spacings.
Details of the lattice discretization and associated
methods are described in Ref.~\cite{Bali:2014kia}. 

The lattice QCD results were obtained using a renormalization point
$\mu=\LambdaH$, where $\LambdaH=120(9) \textmd{ MeV}$ is a
non-perturbatively determined hadronic scale defined by
the condition that at $T=0$ there be no
$\O(B^2)$ contribution to the matter free energy.
In other words, the total free energy
(the sum of matter and magnetic contributions)
equals $\half B^2/e^2(\LambdaH) + \O(B^4)$
in the zero temperature limit.
For nonzero temperatures,
the $\O(B^2)$ contribution to the matter free energy becomes 
nonzero, i.e., the system develops a non-trivial magnetic permeability.

We begin the discussion with the anisotropic pressure components $p_T$ and $p_L$.
To facilitate a comparison with SYM theory, it is natural to 
consider the dimensionless ratio
\begin{equation}
    R\equiv p_T/p_L \,.
\end{equation}
This combination was shown as a function of the temperature
for various values of magnetic field $B$ in the left panel of 
Fig.~\ref{fig:scaling}. 
Notice that the longitudinal pressure $p_L$ is always positive,
so the ratio $R$ remains finite for all $T$ and $B$. 
For low magnetic fields the ratio $R\approx 1$, signaling the near-isotropy of the 
system.
As the field $B$ grows the anisotropy becomes more pronounced and the ratio
$R$ shifts away from unity --- in fact it becomes negative when the transverse
pressure $p_T$ changes sign (and becomes a ``suction'') for strong magnetic fields~\cite{Bali:2014kia}.

A remarkable feature of the results for $R(B,T,\LambdaH)$ is their near-universal
nature when expressed in terms of the dimensionless variable $T/\sqrt{B}$.
This is shown in the right panel Fig.~\ref{fig:scaling},
which plots the same data with the exception of the $B=0$ set. 
The data from different values of the magnetic field all collapse
onto a single curve.
(As noted earlier, small deviations from this scaling behavior
may be present at the lowest temperatures and highest magnetic fields,
but the growth of error bars in this regime prevents any
definitive statement.)
This indicates an apparent universality analogous to what one would
expect, a-priori, only in conformal theories.
In QCD the ratio $R$ is in general a function of two independent
dimensionless parameters,%
\footnote
    {%
    We neglect to indicate explicitly
    additional dependence on the ratios of quark masses to $\LambdaH$.
    }
\begin{equation}
    R(B,T,\LambdaH) = r(T/\sqrt{B}, \LambdaH/\sqrt{B}) \,.
\end{equation}
The apparent near absence 
of any significant dependence on $\LambdaH/\sqrt B$
motivates us to compare the magnetoresponse of QCD to that of
conformal SYM theory, for which $T/\sqrt B$ is the only relevant
dimensionless ratio.

\begin{figure}
\includegraphics[height=2.6in]{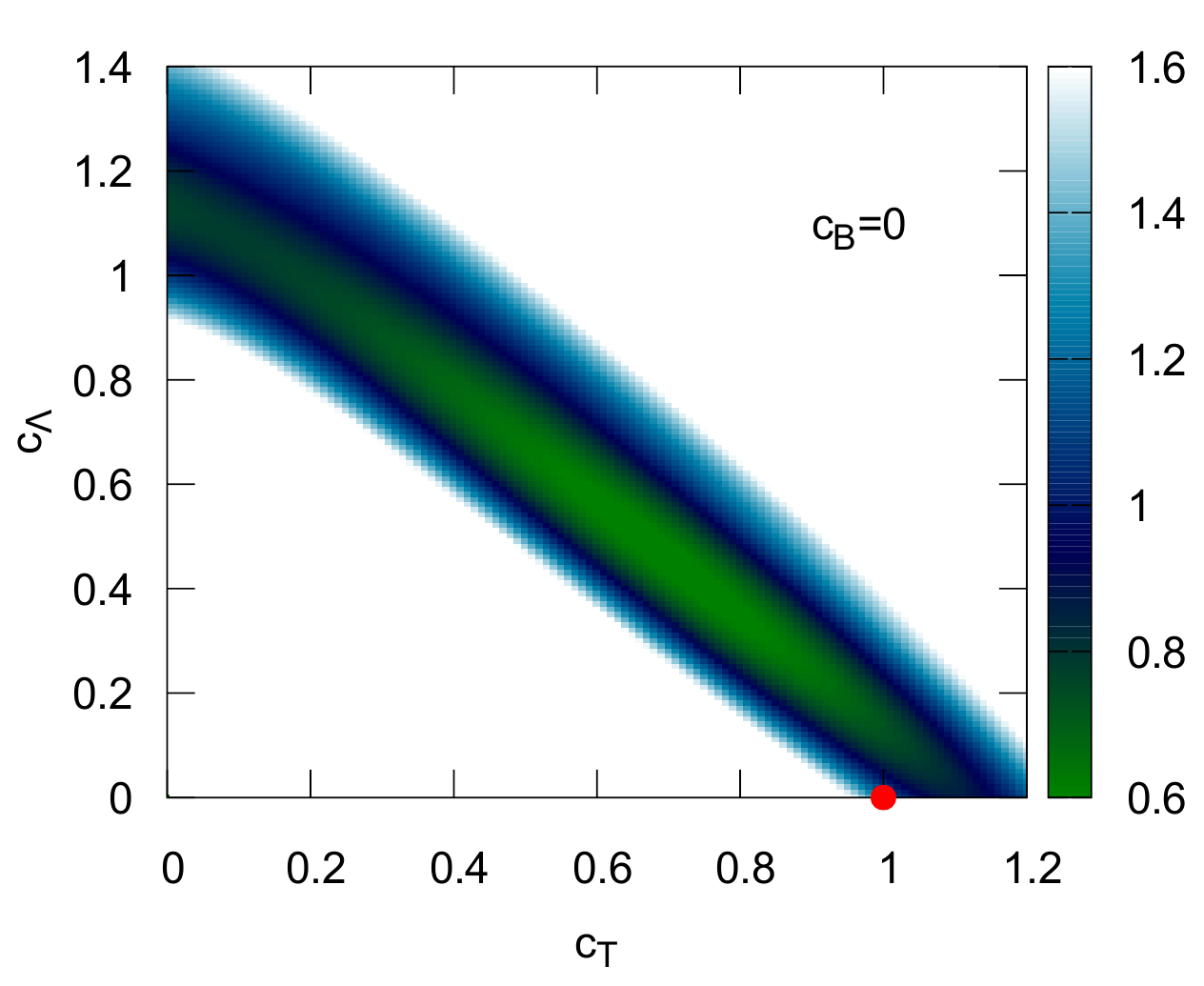}
\hfill
\includegraphics[height=2.6in]{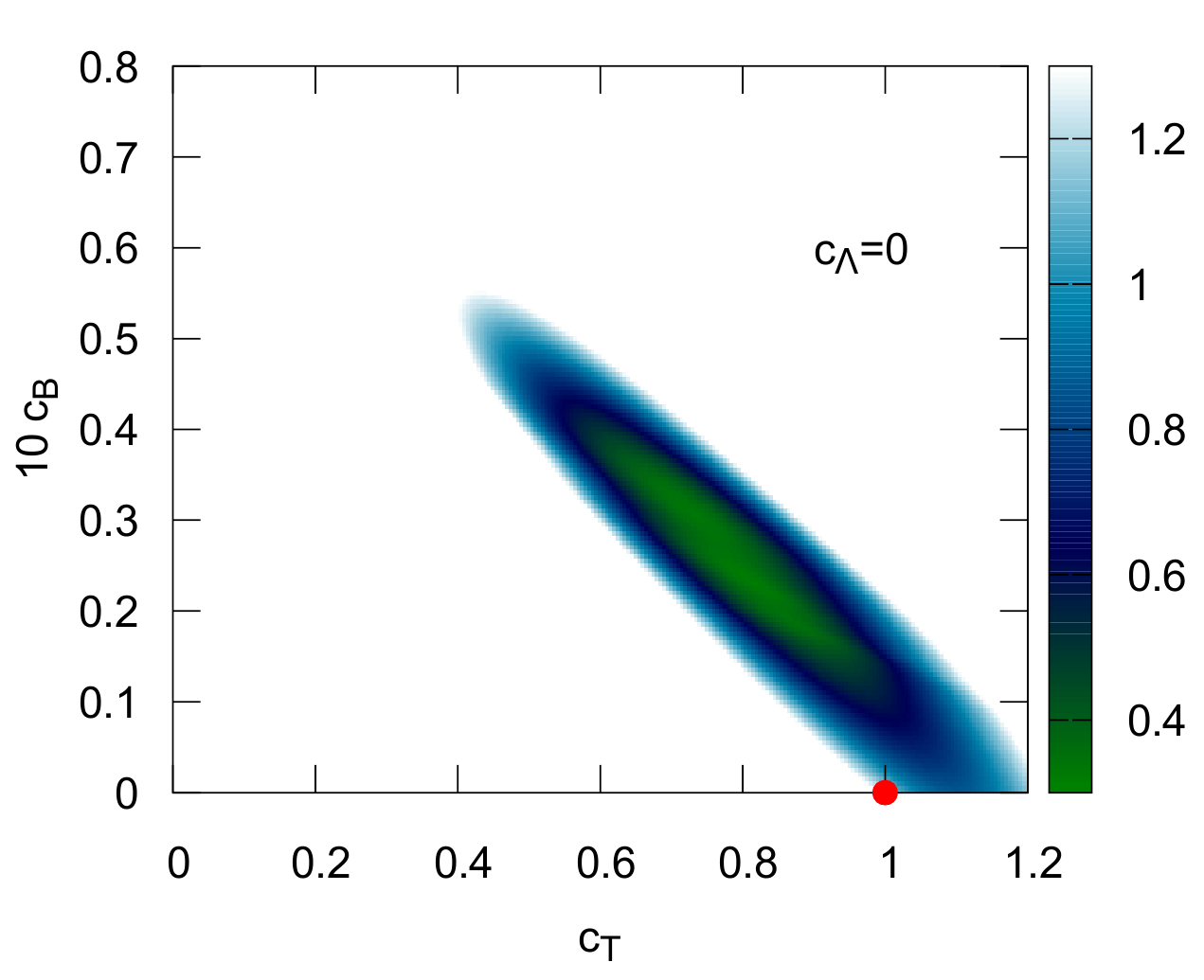}
\caption{\label{fig:deviation}
    Heat map of the deviation from universality, $D$,
    as a function of $c_T$ and $c_\Lambda$ for fixed $c_B = 0$ (left panel), and
    as a function of $c_T$ and $c_B$ for fixed $c_\Lambda = 0$ (right panel).
    The red dot indicates the choice of $\mu=T$.
    The uncertainty in the value of $D$ is of order unity.
    Inside the dark colored regions, where $D \lesssim 1$, 
    universality holds to within the error bars of the lattice data.}
\end{figure}

An important question is how the near-universality of $p_T/p_L$ is affected by 
a change in the renormalization point. We consider a general choice,
\begin{equation}
 \mu(c_T,c_\Lambda,c_B) \equiv
 \sqrt{c_T T^2 + c_B |B| + c_\Lambda \Lambda^2_{\rm H} }\,,
 \label{eq:renscale}
\end{equation}
involving the three underlying scales \{$T$, $B$, $\LambdaH$\}
characterizing the equilibrium state,
parameterized by three coefficients $c_T$, $c_B$ and $c_\Lambda$.
For a quantitative description we introduce a measure $D$ of the
deviation from universality,
\begin{equation}\label{eq:D}
 D\equiv
 \frac{1}{N}\sum_{b,b'} \sum_t
 \frac{[r(t,b)-r(t,b')]^2}{\sigma^2(t,b)+\sigma^2(t,b')} \,,
 \qquad t\equiv\frac{T}{\sqrt{B}}\,,
 \qquad b\equiv\frac{\LambdaH}{\sqrt{B}}\,,
\end{equation}
where $\sigma$ denotes the error of the ratio $r$ and 
the sum extends over all points available from the lattice study
of Ref.~\cite{Bali:2014kia}.
The integer $N$ counts the number of terms in
the resulting sums. 
With this normalization, $D\lesssim 1$ indicates that the curves
for different magnetic fields all overlap each other within errors.
Additionally, the inherent uncertainty on $D$ is of order unity. 
The deviation $D$ is plotted in Fig.~\ref{fig:deviation} as a
heat map in the space of the coefficients $c_B$, $c_T$ and $c_\Lambda$.
The left panel of the figure shows the $c_B=0$ slice, 
while the right panel shows the orthogonal $c_\Lambda=0$ slice.%
\footnote
    {%
    The heat map of $D$ remains very similar if one instead uses a
    parameterization of the renormalization point
    which is analytic in the magnetic field,
    $\mu=\big(c_T T^4+ c_B B^2+ c_\Lambda \LambdaH^4\big)^{1/4}$.
    } 

The ansatz \eqref{eq:renscale}, reflecting the presence of three
different potentially relevant underlying scales, is a natural form
for parameterizing a dominantly relevant scale in QCD.
The appearance of $\LambdaH$ reflects the intrinsic lack of
scale symmetry in QCD (even in the limit of massless quarks).
In a conformal theory, there is no intrinsic energy scale, and hence
no equivalent of the QCD scale $\LambdaH$.
A natural question to ask is whether the apparent universality in our
lattice data remains evident when $c_\Lambda$ is small or zero.
As indicated in the left panel of Fig.~\ref{fig:deviation},
this is indeed the case.
The region of minimal deviation from universality is 
found to be centered around the point $(c_T,c_\Lambda)=(0.70,0.46)$,
but it extends out to include, for example,
the purely temperature-driven renormalization point $\mu=T$.

Therefore, in the following we set $c_\Lambda$ to zero, so that a comparison to 
$\Nfour$ SYM theory will be straightforward. 
As shown in the right panel of Fig.~\ref{fig:deviation},
for vanishing $c_\Lambda$ the minimum of $D$ defines a
valley along the line $c_B=0.087-0.084 \,c_T$.
Below we will compare QCD to the SYM theory along this valley.

\begin{figure}
\includegraphics[height=2.6in]{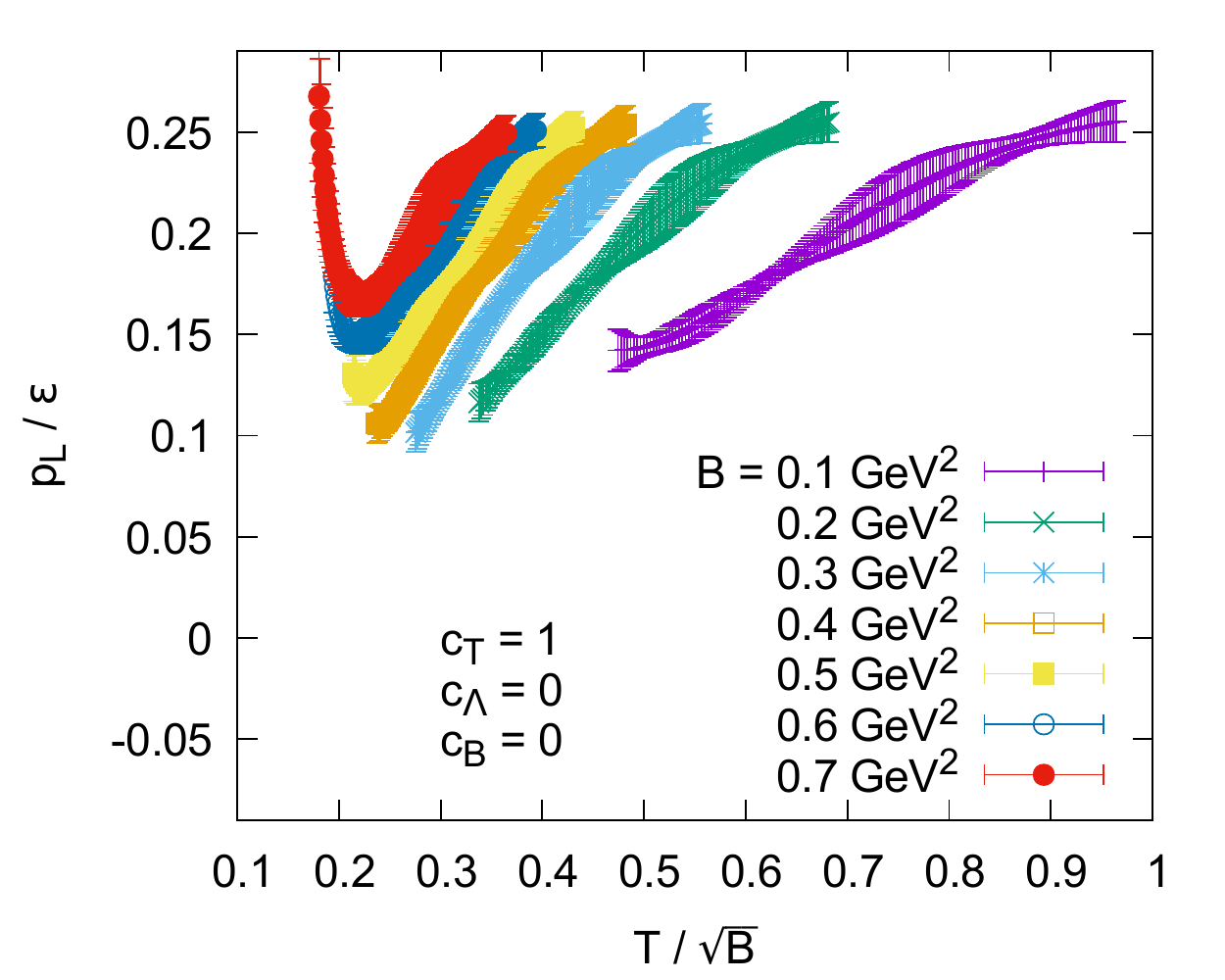}\!\!\!\!
\includegraphics[height=2.6in]{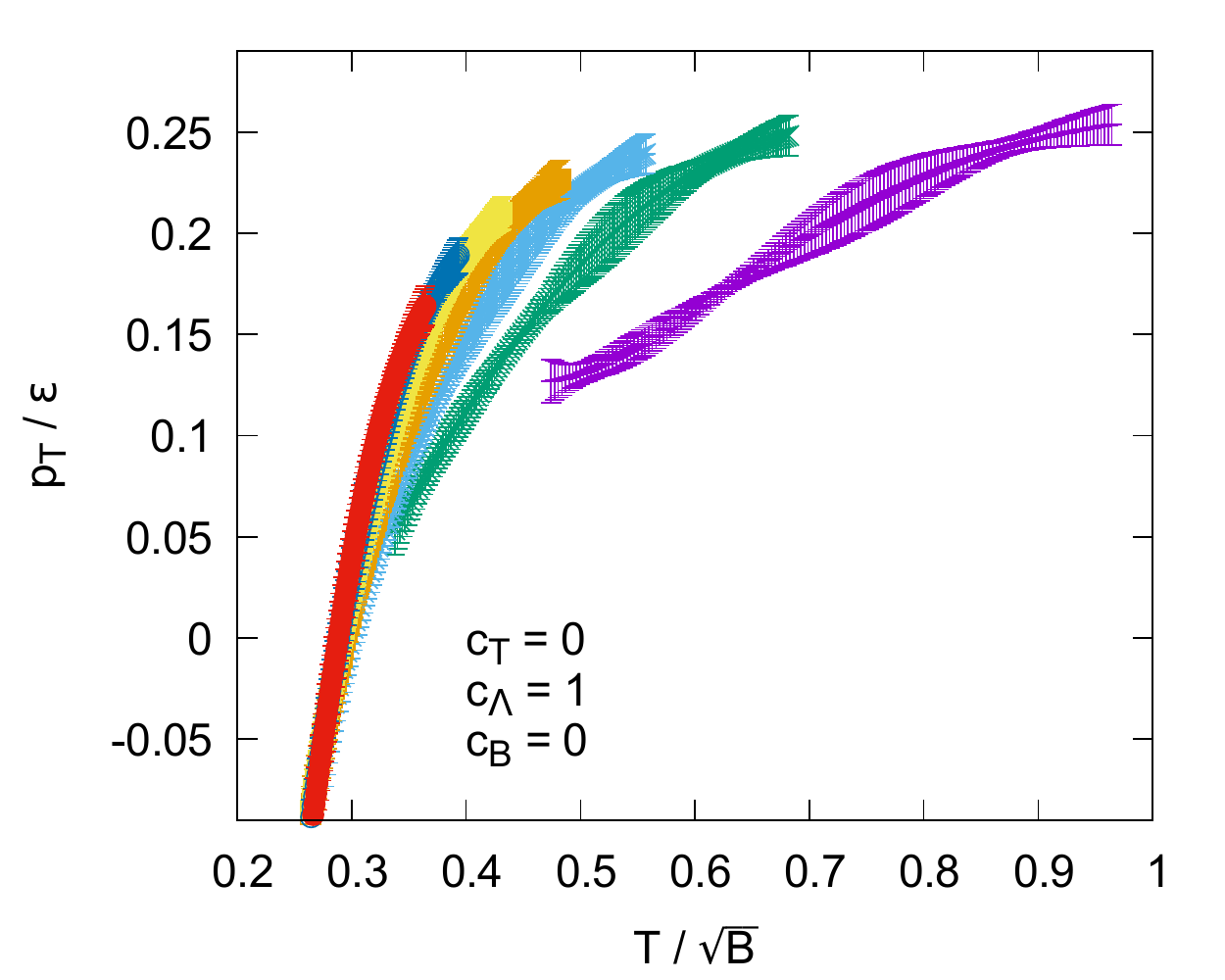}
\caption{\label{fig:nonUniversality}
   The ratio $p_L/\epsilon$ at the renormalization point $\mu=T$ (left panel) and 
   $p_T/\epsilon$ at $\mu=\Lambda_H$ (right panel) for various values of the magnetic field $B$, 
    and plotted as a function of $T / \sqrt{B}$. 
    Unlike the ratio $p_T/p_L$, no near universal behavior is observed
    in either ratio involving the energy density.}
\end{figure}

As illustrated in Fig.~\ref{fig:nonUniversality}, 
other dimensionless ratios, such as $p_L/\epsilon$ or $p_T/\epsilon$,
have substantial dependence on $\LambdaH/\sqrt B$ and do not
exhibit the near universality seen in the pressure anisotropy ratio $p_T/p_L$.
Given the connection \eqref{eq:I} between the trace anomaly and the
interaction measure $I = \epsilon -2 p_T -p_L$,
this surely reflects the substantial growth of the interaction measure in QCD
as the temperature approaches the confinement transition
due to the intrinsic violation of scale invariance in QCD
\cite{Borsanyi:2013bia,Bazavov:2014pvz},
a feature not present in conformal $\Nfour$ SYM.

\section{$\mathcal{N}=4$ supersymmetric Yang Mills theory} \label{sec:SYM}

We consider maximally supersymmetric ($\Nfour$) $SU(\Nc)$ Yang-Mills theory,
in the limit of large $\Nc$ and large 't Hooft coupling, $\lambda \gg 1$, 
coupled to a background ``electromagnetic'' $U(1)$ gauge field.
To define this coupling, we choose the same $U(1)$ subgroup of the
$SU(4)$ global $R$-symmetry which was used in Ref.~\cite{Fuini:2015hba}.
Specifically,
\begin{equation}
    q_w^\alpha = (3,-1,-1,-1)/\sqrt 3 \,,\qquad
    q_s^a = (2,2,2)/\sqrt 3 \,,
\label{eq:SYMcharges}
\end{equation}
are the respective charge assignments
for the four Weyl fermions and three complex scalars of $\Nfour$ SYM.
With these assignments, the $U(1)$ $\beta$-function coefficient
(\ref{eq:EMbeta_SYM}) becomes
\begin{equation}
    b_1 = \frac{\Nc^2{-}1}{4\pi^2} \,.
    \qquad\qquad\mbox{[$\Nfour$ SYM]}
\label{eq:SYMb_1}
\end{equation}
(However, as discussed in the next section, when comparing with QCD
we will rescale the above charge
assignments by an adjustable factor.)

Equilibrium states of this theory, in the presence of a homogeneous background
magnetic field, have a dual gravitational description
given by magnetic black brane solutions
first computed by D'Hoker and Kraus~\cite{D'Hoker:2009mm}.
These are solutions of 5D Einstein-Maxwell theory, which is a consistent
truncation of type IIB supergravity.
The Einstein-Maxwell action
\begin{equation} \label{eq:EMAction}
    S = \frac{1}{16\pi G_5} \int d^5 x \, \sqrt{-g} \,
    \left [
	\mathcal{R}-2\Lambda - L^2 \, \mathcal{F}_{MN} \mathcal{F}^{MN}
    \right ]
    + \theta \int d^5 x \> \mathcal{A}\wedge \mathcal{F} \wedge \mathcal{F} \,,
\end{equation}
where $M,\, N = 0,\,{\ldots},4$ are the 5D spacetime indices, 
$g$ is the metric, $\mathcal{R}$ is the Ricci scalar,
$G_5 = \frac \pi{2} L^3 / (\Nc^2{-}1)$ is the 5D 
Newton gravitational constant,
$\Lambda = -6 L^{-2}$ is the cosmological constant and
$\mathcal{F}=d\mathcal{A}$ is the five-dimensional electromagnetic
field strength.
The reduction from IIB supergravity leads to a specific value for
the Chern-Simons coupling $\theta$, but this term vanishes identically
for our solutions of interest and may be ignored.
Solutions to the gravitational theory (\ref{eq:EMAction})
representing uncharged (magnetic)
black branes may be described by a metric of the form
\cite{D'Hoker:2009mm},
\begin{equation} \label{eq:magneticBlackBraneAnsatz}
    ds^2 = 
    -U(r) \, dt^2 + \frac{dr^2}{U(r)}
    +e^{2 V(r)} \, ({dx}^2+{dy}^2)
    +e^{2 W(r)} \, dz^2  \,,
\end{equation}
plus a bulk field strength
\begin{equation}
    \F = \B \, dx \wedge dy   \,,
\end{equation}
representing a constant magnetic field of strength $B$.
The metric functions $U$, $V$ and $W$ depend only on the radial coordinate $r$,
and must be computed numerically.
These functions have the near-boundary ($r \to \infty$) asymptotic behavior,
\begin{subequations}
\begin{align}
    U(r) &= r^2 /L^{2}
    + 2 (a_4 - \tfrac 13 \B^2 \ln r/L) \, L^6 / r^{2}
    + \O(r^{-6} \, \ln^2 r/L) \,,
\\
    V(r) &= \ln (r/L)
    + \tfrac 12 (b_4 + \tfrac 13 \B^2 \ln r/L) \, L^8 /r^4
    + \O(r^{-8} \, \ln^2 r/L) \,,
\\
    W(r) &= \ln (r/L)
    - (b_4 + \tfrac 13 \B^2 \ln r/L) \, L^8 /r^4
    + \O(r^{-8} \, \ln^2 r/L) \,.
\end{align}
\end{subequations}
The subleading terms in these near-boundary expansions determine
the (expectation value of the) $\Nfour$ SYM stress-energy tensor
\cite{Fuini:2015hba}.
Specifically,
\begin{subequations}
\begin{align}
    \langle T^{tt}\rangle &= \epsilon
    = \kappa\left(-\tfrac{3}{2} \, a_4 + \half\B^2 \, \ln\mu L\right) ,
\\
    \langle T^{xx}\rangle = \langle T^{yy}\rangle
    &= p_T
    = \kappa\left(-\half a_4 + b_4 - \tfrac{1}{4}\B^2 + \half\B^2\ln\mu L\right) ,
\\
    \langle T^{zz}\rangle &= p_L
    = \kappa\left(-\half a_4 - 2b_4 - \half\B^2\ln\mu L\right)  ,
\end{align}
\end{subequations}
with all off-diagonal components vanishing.
Here $\kappa \equiv (\Nc^2{-}1)/(2\pi^2)$ and $\mu$, once again,
is the arbitrary renormalization point used to separate the SYM 
and background EM contributions to the total stress-energy tensor.
For further details, see Refs.~\cite{Fuini:2015hba,Janiszewski:2015ura}.

\section{Comparison of QCD and $\Nfour$ SYM} \label{sec:comparison}

To compare lattice QCD results with thermodynamic data for $\Nfour$ SYM 
calculated via holography, one must decide how best to adjust for the
differing field content of the two theories.
Specifically, in making this comparison should the SYM charge assignments
be rescaled?
The overall normalization of our SYM charge assignments
(\ref{eq:SYMcharges}) was merely a convenient choice
which corresponds to the absence of additional numerical factors multiplying
the Maxwell term in the dual gravitational action (\ref{eq:EMAction})
\cite{Fuini:2015hba}.
A uniform rescaling of charge assignments is equivalent to a rescaling of
the magnetic field, so this question is the same as asking whether 
comparisons are most usefully made at coinciding values of 
magnetic field, as it was introduced in the gravitational action
(\ref{eq:EMAction}),
or whether it is appropriate to first rescale the background
magnetic field added to SYM theory.

As long as the background electromagnetic field is treated as classical,
the normalization of SYM charge assignments is arbitrary, as there is no
intrinsic scale available to define physical units in which to measure a
magnetic field.
In other words, there is no quantization of EM charges or magnetic fluxes.
Hence, it is completely appropriate to rescale SYM charge assignments,
or equivalently rescale the magnetic field, $B \to B/\xi$ for some choice
of $\xi$, when comparing with QCD.
(In contrast, temperature may be regarded as having a common operational
meaning in both theories, so no rescaling of temperature is performed.)
The key question is how should one choose this charge (or magnetic field)
scale factor $\xi$?

From our earlier discussion (sec.~\ref{sec:renorm}) of renormalization
point dependence,
one seemingly natural possibility to consider is scaling the SYM charge
assignments so that the leading coefficient $b_1$ in the $U(1)$
$\beta$-function (\ref{eq:b1}) coincides between QCD and SYM.
This would require scaling the SYM charge assignments inversely with
$\Nc$ so as to compensate for the difference in the number of charged
degrees of freedom.
However, this choice is neither necessary nor helpful,
as one can always first define rescaled stress-energy tensors,
$
    \widetilde T^{\mu\mu} \equiv T^{\mu\nu}/b_1
$,
in both QCD and SYM,
so that the rescaled tensors satisfy identical renormalization
group equations.
Since we are comparing dimensionless ratios such as $p_T/p_L$,
such an overall rescaling of $T^{\mu\nu}$ has no effect on the
comparison between theories.

Another possible approach involves matching the magnetoresponse in the
asymptotically high temperature limit.
As discussed in appendix \ref{sec:appB}, if one considers the entropy
density (which is independent of the renormalization point $\mu$) and
demands that the relative contribution of the $\O(B^2)$ terms coincide,
so that
\begin{equation}
    \left. \frac {s(B,T)}{s(0,T)} \right|_{\rm QCD}
    =
    \left. \frac {s(B/\xi,T)}{s(0,T)} \right|_{\rm SYM}
    + \O (B^4) \,,
\end{equation}
then this condition leads to a scale factor choice
$
    \xi = \sqrt{19/3} \approx 2.5
$.

This value for the charge rescaling defines an arguably sensible
scheme for comparing our two theories.
However, it uses information from asymptotically high temperature QCD
which is far from the regime of a few times $T_{\rm c}$ where it is
appropriate to view real quark-gluon plasma as a strongly coupled
near-conformal fluid.
Consequently, our preferred approach is the simplest: we just treat
the charge rescaling factor $\xi$ as a free parameter, and find the value
which minimizes the difference between the QCD and SYM results for
the pressure ratio $R = p_T/p_L$.
More precisely, 
we first evaluate this pressure ratio (as a function of $B$ and $T$),
in both theories at a common renormalization point
$\mu = \left( c_T T^2 + c_B |B| \right)^{1/2}$
which lies along the valley defined by $c_B =0.087-0.084 \,c_T$.
As shown in Fig.~\ref{fig:deviation},
along this valley the QCD ratio $R$ is essentially a function of only
the single variable $T/\sqrt{B}$.
We then define a deviation $\Delta R$ between the QCD pressure ratio and
that of SYM,
\begin{equation}\label{eq:deltaR}
    \Delta R (\xi, c_T)
    \equiv 
    \frac{1}{N}  \sum_{B,T} \>
    \frac{\Big [ R^{\rm QCD}(T/\sqrt B;\mu(c_T))
		- R^{\rm SYM}(T \sqrt{\xi/B};\mu(c_T)) \Big]^2}{\sigma^2(T,B)} ,
\end{equation}
with the SYM magnetic field rescaled by a adjustable factor $\xi$
and $c_B$ determined by $c_T$ along the aforementioned valley.
Only lattice QCD data points with $T > 150$ MeV are included in this sum,
as lower temperatures probe the hadronic phase of QCD, not the deconfined
plasma phase.
As in our earlier measure $D$ of the deviation from universality \eqref{eq:D},
$\sigma$ denotes the lattice error of the ratio $R$, and
$N$ is the number of terms in the sum.
A value less than unity for the deviation, $\Delta R<1$,
indicates agreement between the two theories to within the errors of the
lattice results, and the inherent uncertainty in $\Delta R$ is of order unity.

\begin{figure}[htp]
\includegraphics[height=3.0in]{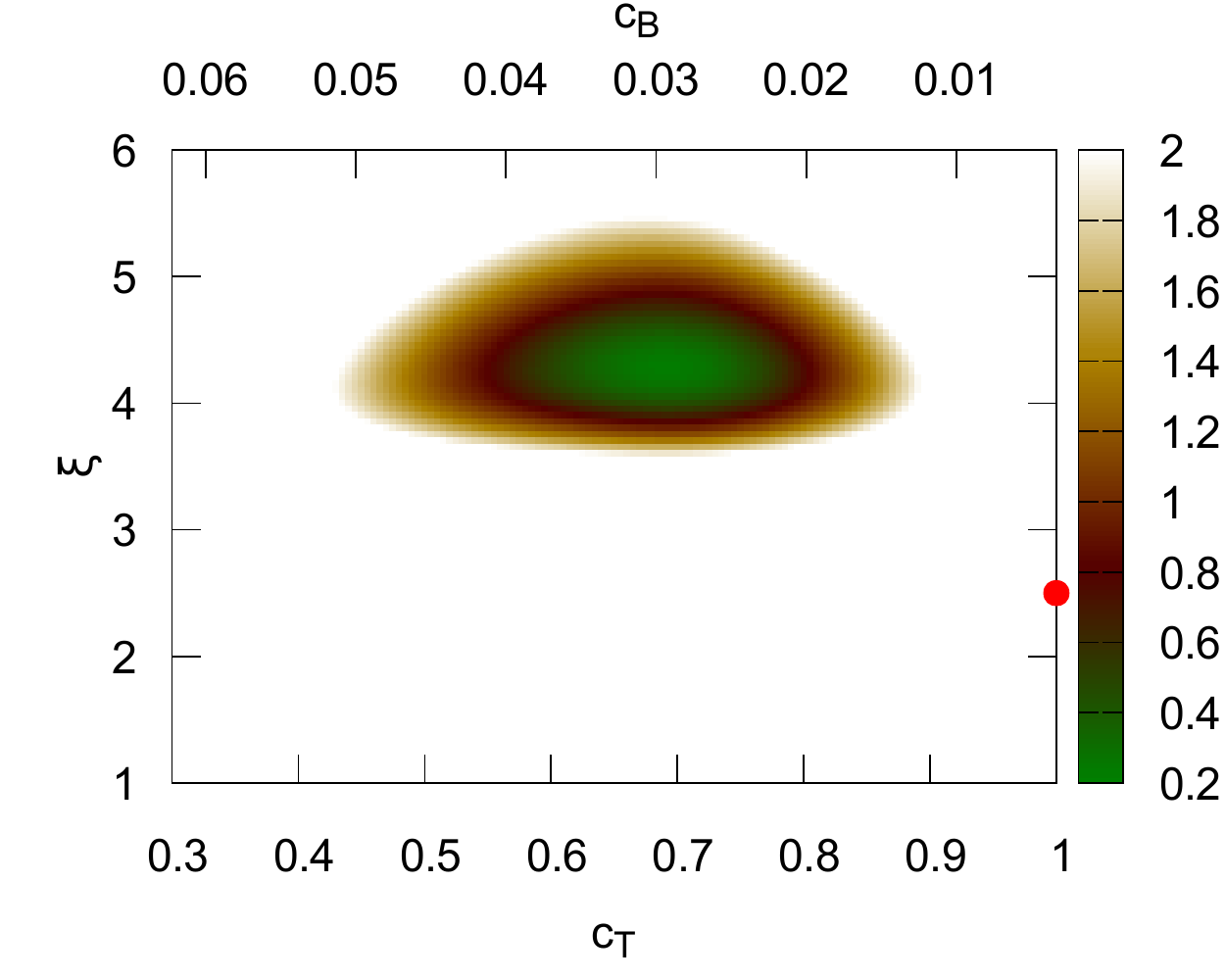}
\caption{\label{fig:valleyAm} 
The normalized deviation $\Delta R$ of the QCD and $\Nfour$ SYM results for
$p_T/p_L$, plotted as a function of the SYM charge rescaling factor $\xi$
and the value of $c_T$ defining the renormalization point
(with $c_B$ correspondingly fixed to lie along the QCD valley of near-universality).
The red dot indicates the high temperature motivated choice
$\mu = T$ and $\xi = 2.5$.}
\end{figure}

Our results for $\Delta R$ are plotted in Fig.~\ref{fig:valleyAm}.
As clearly seen in the figure, $\Delta R$ develops a minimum
around $\xi=4.3$ and $c_T=0.69$ (implying $c_B=0.029$),
with its minimum value well below 1.
The red dot on the right of Fig.~\ref{fig:valleyAm} indicates a choice 
of the high temperature motivated rescaling factor
$\xi = 2.5$ discussed above combined with $c_T = 1$.
At this point $\Delta R$ is large compared to 1,
indicating much less satisfactory matching
between theories with this choice of rescaling.

\begin{figure}[htp]
\includegraphics[height=3.0in]{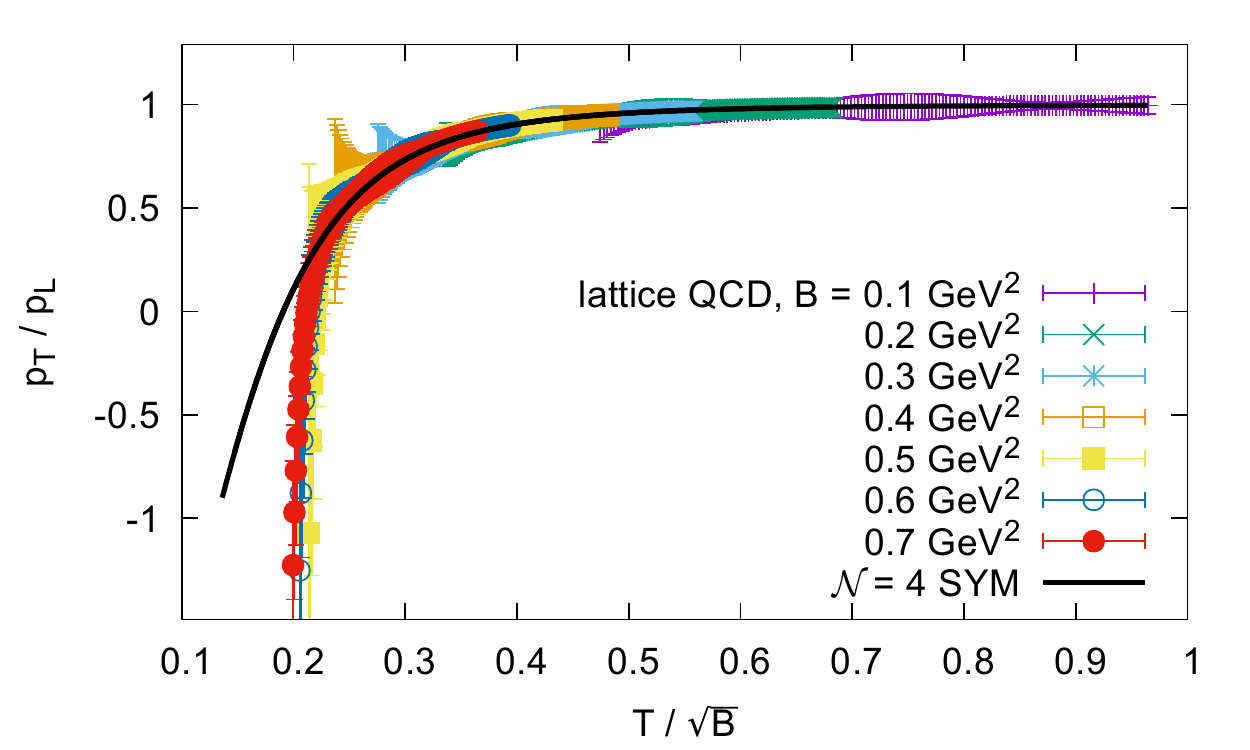}
\caption{\label{fig:optimizedPressureRatio}
The ratio $p_T/p_L$ from lattice QCD as a function of $T/\sqrt{B}$ at optimal universality -- i.e.\ for a renormalization scale
parameterized by $c_B=0.029$, $c_T=0.69$ and $c_\Lambda=0$. 
Also included is the holographic pressure ratio 
computed at the same renormalization scale 
and with the electric charge normalization factor $\xi = 4.3$.
}
\end{figure}

A direct comparison of the pressure anisotropy ratio $p_T/p_L$ in
QCD and $\Nfour$ SYM is displayed in Fig.~\ref{fig:optimizedPressureRatio}
using choices of renormalization point and charge rescaling which
minimize $\Delta R$, namely
$(c_T,c_B)=(0.69,0.029)$ (with $c_\Lambda=0$),
and $\xi=4.3$.
In addition to the universal scaling of the lattice QCD data,
one sees that the SYM curve lies atop the error bars of the
QCD data for all $T/\sqrt B \gtrsim 0.22$, or equivalently
for magnetic fields up to $\approx 21 \, T^2$.
Deviations of the QCD data from the SYM curve are present,
and are significant, for $T/\sqrt{B} \lesssim 0.22$.
This reflects the limit of the region where it makes
sense to model QCD plasma as a conformal fluid.
Figure~\ref{fig:optimizedPressureRatioDiag}, shown in the Introduction,
gives a global view of the region of the temperature-magnetic field plane,
in physical units, in which agreement between the QCD and SYM magnetoresponse
holds to within the error estimate on the lattice QCD value of $p_T/p_L$.
Agreement was inevitable at large $T/\sqrt B$ where
the pressure ratio in both theories necessarily approaches unity.
But excellent agreement down to rather small temperatures,
or up to quite large magnetic fields where the deviation of the
pressure anisotropy from unity is substantial, is surprising.
More precisely, it is remarkable that a choice of renormalization scale exists
for which the pressure ratio in QCD and SYM, suitably compared,
displays a common conformal behavior over such a substantial range
of temperature and magnetic field.

\section{Discussion} \label{sec:discussion}

In this work, we analyzed data from a recent lattice gauge theory 
calculation of the thermodynamics of a QCD plasma placed in
an external magnetic field.
Except at asymptotically high temperatures, $T \gggtr \Lambda_{\rm QCD}$,
observables in QCD will generically have independent non-trivial dependence
on the value of both temperature and magnetic field
(relative to $\Lambda_{\rm QCD}$).
Moreover,
separating the QCD contribution from the classical Maxwell contribution
to the stress-energy tensor necessarily introduces dependence on an arbitrary
renormalization point $\mu$, as discussed in section~\ref{sec:renorm}.
So in the presence of a non-zero magnetic field, at any physically
accessible temperature,
the transverse to longitudinal pressure ratio $R = p_T/p_L$
should be expected to display non-trivial dependence on multiple
dimensionless ratios, for example $T/\sqrt B$, $T/\Lambda_{\rm QCD}$,
and $T/\mu$.
Choosing the renormalization point to depend in some dimensionally
consistent fashion on the physical scales $T$, $\sqrt B$,
and $\Lambda_{\rm QCD}$ still leaves two independent dimensionless
ratios on which the pressure ratio should depend.
However,
as shown in Fig.~\ref{fig:scaling},
we find that for suitable choices of renormalization point
the pressure anisotropy $p_T/p_L$ exhibits scale invariance
to within the error estimates of the lattice data,
with functional dependence only on the ratio $T/\sqrt B$.
A more careful study of the deviations from universality identified
an optimal choice of renormalization point,
$\mu=\left({0.69 T^2+ 0.029 |B|}\right)^{1/2}$
(given our specific measure (\ref{eq:D}) on the deviation).

Scale invariance is, of course, a feature of conformal field theories.
Our observed near-perfect scale invariance in the QCD pressure anisotropy
motivated a comparison with the pressure anisotropy in the
simplest four dimensional conformal gauge theory, 
$\Nfour$ supersymmetric Yang-Mills (SYM) theory,
when this theory is placed in an external magnetic field.
Specifically, we compared with $\Nfour$ SYM in the large $\Nc$
and strong coupling limit, for which a dual gravitational description
is available.
After a simple matching of the electromagnetic couplings of the 
two theories, the SYM pressure anisotropy was found to agree with that
of QCD over a wide range of temperature and magnetic field values,
as shown in Fig.~\ref{fig:optimizedPressureRatio}.
This agreement persists at unexpectedly low temperatures and
large magnetic fields.
(Growing error bars on the lattice data make the comparison
inconclusive below $T/\sqrt{B}<0.2$.)
The region where the pressure
ratios of the two theories coincide was visualized in
Fig.~\ref{fig:optimizedPressureRatioDiag}.
It must be noted, however, that ratios of other thermodynamic quantities
do not exhibit the same universal scaling behavior seen in
the pressure anisotropy $p_T/p_L$.
For ratios involving the energy density such as $p_T/\epsilon$ or
$p_L/\epsilon$, no choice of renormalization scale creates an overlap 
of data from different values of external field anywhere near as
striking as that seen in the pressure anisotropy ratio.
This, presumably, reflects the substantial peak in the thermal
expectation value of the QCD trace anomaly, $I = \epsilon - 2 p_T - p_L$,
near the QCD confinement transition, which is not reproduced by
$\Nfour$ SYM.

Although our analysis has exclusively involved equilibrium quantities
(for which lattice QCD calculations are possible), the region of
``effective conformality'' shown in Fig.~\ref{fig:optimizedPressureRatioDiag}
is presumably also the region in which the long wavelength dynamics of
QCD plasma is reasonably well described by conformal hydrodynamics.
Outside this region, effects of scale non-invariance should be
increasingly important, implying significant bulk viscosity effects
in QCD hydrodynamic response.

This work adds the pressure anisotropy magnetoresponse, 
described by a non-trivial scaling function of $T/\sqrt B$, to the
set of thermal observables in QCD which are well-reproduced by
strongly coupled $\Nfour$ SYM, the simplest (four dimensional)
conformal gauge theory with a holographic description.
It also reveals the limitations of modeling hot QCD plasma as
a conformal fluid when thermodynamic ratios
involving the energy density are examined.
This limitation is unsurprising, given what is known about
the temperature dependence of the trace anomaly expectation value.

It would be interesting to explore extensions of this work 
involving comparisons with other strongly coupled theories
having holographic descriptions which are closer to QCD
than $\Nfour$ SYM.
Possibilities include $\mathcal N\,{=}\,2^*$ SYM 
\cite {Donagi:1995cf,
    Pilch:2000ue,
    Buchel:2003ah,
    Hoyos:2011uh}
and other mass deformations of $\Nfour$ SYM,
cascading gauge theory \cite{Klebanov:2000hb,Aharony:2007vg},
the Sakai-Sugimoto model \cite{Isono:2015uda,Sakai:2004cn},
and various bottom-up models
(for example, \cite{Attems:2016ugt}).
Turning on additional deformations which can be studied
both in lattice QCD and in strongly coupled holographic models,
such as a non-zero isospin chemical potential,
could also be instructive.
Such work is left for the future.

\acknowledgments
    {%
    This work was supported, in part,
    by the DFG (Emmy Noether Programme EN 1064/2-1) and
    by the U.~S.~Department of Energy grants DE-SC-0012447 and
    DE-SC\-0011637.
    LY gratefully acknowledges the hospitality of the University of Regensburg 
    and generous support from the Alexander von Humboldt foundation.
    }

\appendix

\section{Magnetic field matching at high temperature}\label{sec:appB}

To define an optimal matching of the high temperature magnetoresponse in
QCD and $\Nfour$ SYM, we focus on the entropy density
$s = -\partial f / \partial T |_\mu$,
as this quantity is independent of the choice of renormalization point $\mu$.
As discussed in Ref.~\cite{Bali:2014kia},
at asymptotically high temperature where $T$ is the only relevant physical
scale, the renormalization group equation (\ref{eq:RG-f}) for the QFT
free energy density (namely
$
    \mu \frac d{d\mu} f = b_1 \, B^2
$)
plus dimensional analysis
implies that 
$f = f_0 \, T^4 - b_1 \, B^2 \, [\ln (T/\mu) + \mbox{const.}] + \O(B^4)$,
where $f_0 T^4$ is the free energy density at zero magnetic field.
Consequently, the entropy density has the form
\begin{equation}
    s = s_0 + b_1 \, B^2/T + \O(B^4) \,.
\end{equation}

For three flavor QCD at temperatures $T \gg \LambdaH$, asymptotic freedom
implies that the entropy density approaches the Stefan-Boltzmann limit,
so $s_0 = \frac{19}{9} \pi^2 T^3$.
And from Eq.~(\ref{eq:EMbeta_QCD}),
the EM $\beta$-function coefficient $b_1 = 1/(6\pi^2)$.
Hence,%
\footnote
    {%
    There is a typo in the expression for $s/T^3$ in Eq.~(4.9)
    of Ref.~\cite{Bali:2014kia}, which is corrected here.
    }
\begin{subequations}
\begin{equation}
    \frac s{T^3}
    =
    \frac{19\pi^2}{9} + \frac {B^2}{6\pi^2 \, T^4} + \O(B^4) \,.
    \hspace*{1.5in}\mbox{[QCD]}
\end{equation}
For $\Nfour$ SYM at strong coupling, the zero field entropy density
$s_0 = (\Nc^2{-}1) \frac{\pi^2}{2} \, T^3$ and,
with the charge assignments (\ref{eq:SYMcharges}),
the $U(1)$ $\beta$-function coefficient $b_1 = (\Nc^2{-}1)/(4\pi^2)$.
If these charge assignments are rescaled by an inverse factor of $\xi$, then
\begin{equation}
    \frac s{T^3}
    =
    (\Nc^2{-}1)
    \left[
    \frac{\pi^2}{2} + \frac {B^2}{4\pi^2 \, \xi^2 \, T^4} + \O(B^4)
    \right] .
    \qquad\mbox{[$\Nfour$ SYM]}
\end{equation}
\end{subequations}
Matching the relative contribution of the $\O(B^2)$ term in the
entropy density,
i.e., demanding that $s(B,T)/s(0,T)$ coincide up to $\O(B^4)$, leads
to $\xi^2 = 19/3$, or $\xi \approx 2.5$.

\bibliographystyle{JHEP} 
\bibliography{lattholo}

\end{document}